\documentclass[aps,prb,twocolumn,showpacs,superscriptaddress]{revtex4-1}  
\usepackage{amsmath}
\usepackage{amssymb}
\usepackage{graphicx}
\usepackage{epstopdf}
\usepackage{suffix}
\usepackage{mathtools}

\begin{document}

\title{Interfacial spin Seebeck effect in noncollinear magnetic systems}

\author{Benedetta Flebus}
\author{Yaroslav Tserkovnyak}
\address{Department of Physics and Astronomy, University of California, Los Angeles, California 90095, USA}
\author{Gregory A. Fiete}
\address{Department of Physics, The University of Texas at Austin, Austin, TX 78712, USA}
\address{Department of Physics, Massachusetts Institute of Technology, Cambridge, Massachusetts 02139, USA}
\begin{abstract}
While the spin Seebeck effect  (SSE)  has proven to be an invaluable tool for probing spin correlations in collinear magnetic insulators, a theory generalizing the heat-to-spin interconversion in noncollinear magnets is still lacking. Nonetheless, a variety of quantum magnets that are attracting an increasing attention to, e.g., their topological properties, display a noncollinear spin order.  Here, we establish a general framework for thermally-driven spin transport at the interface between a noncollinear magnet and a normal metal. Modeling the interfacial coupling between localized and itinerant magnetic moments via an exchange Hamiltonian, we derive an expression for the spin current, driven by a temperature difference, for an arbitrary noncollinear magnetic order. Our theory reproduces previously obtained results for ferromagnetic and antiferromagnet systems.
\end{abstract}

\pacs{}

%75.76.+j		Spin transport effects
%72.25.Mk	Spin transport through interfaces
%72.20.Pa	Thermoelectric and thermomagnetic effects
%85.75.-d		Magnetoelectronics; spintronics: devices exploiting spin polarized transport or integrated magnetic fields

\maketitle

\section{Introduction}

 Historically, thermoelectric phenomena have  been an invaluable tool for probing correlations in electronic materials.
The interplay between thermal and electric properties has been invoked, for instance, in elucidating the non-Fermi properties of metals through the violation of the Wiedemann-Franz law~[\onlinecite{wakeham}], the non-Abelian nature of fractional quantum Hall states~[\onlinecite{mitali}], and the chiral Majorana edge modes in gapped chiral spin liquids~[\onlinecite{spinliquid}]. 

More recently, spin-caloritronic probes, which rely on the coupling between the electron's spin and heat fluxes, have garnered much attention due to their potential for probing magnetic properties of insulating quantum materials~[\onlinecite{Bauer2010}, \onlinecite{Zutic2004}].  In particular, since its observation in a magnetic insulating system~[\onlinecite{uchida2008}], the SSE, i.e., the generation of a pure spin current in response to a temperature gradient, has been subject to intense scrutiny~[\onlinecite{Uchida2014}]. Since pure spin currents pumped out of a magnetic insulator can be converted, via the inverse Spin Hall effect~[\onlinecite{Saitoh2006, Kimura2007,Miao,Sinova2015}], into a transverse charge current in an adjacent normal metal, the SSE has opened  up new prospects for converting otherwise-wasted heat into charge currents.  Thus, in addition to its promise for probing magnetic quantum materials, the discovery of the SSE has  stimulated the pursuit of new material platforms and functionalities for designing efficient thermoelectric devices. 

\begin{figure}[b!]
\includegraphics[width=0.8\linewidth]{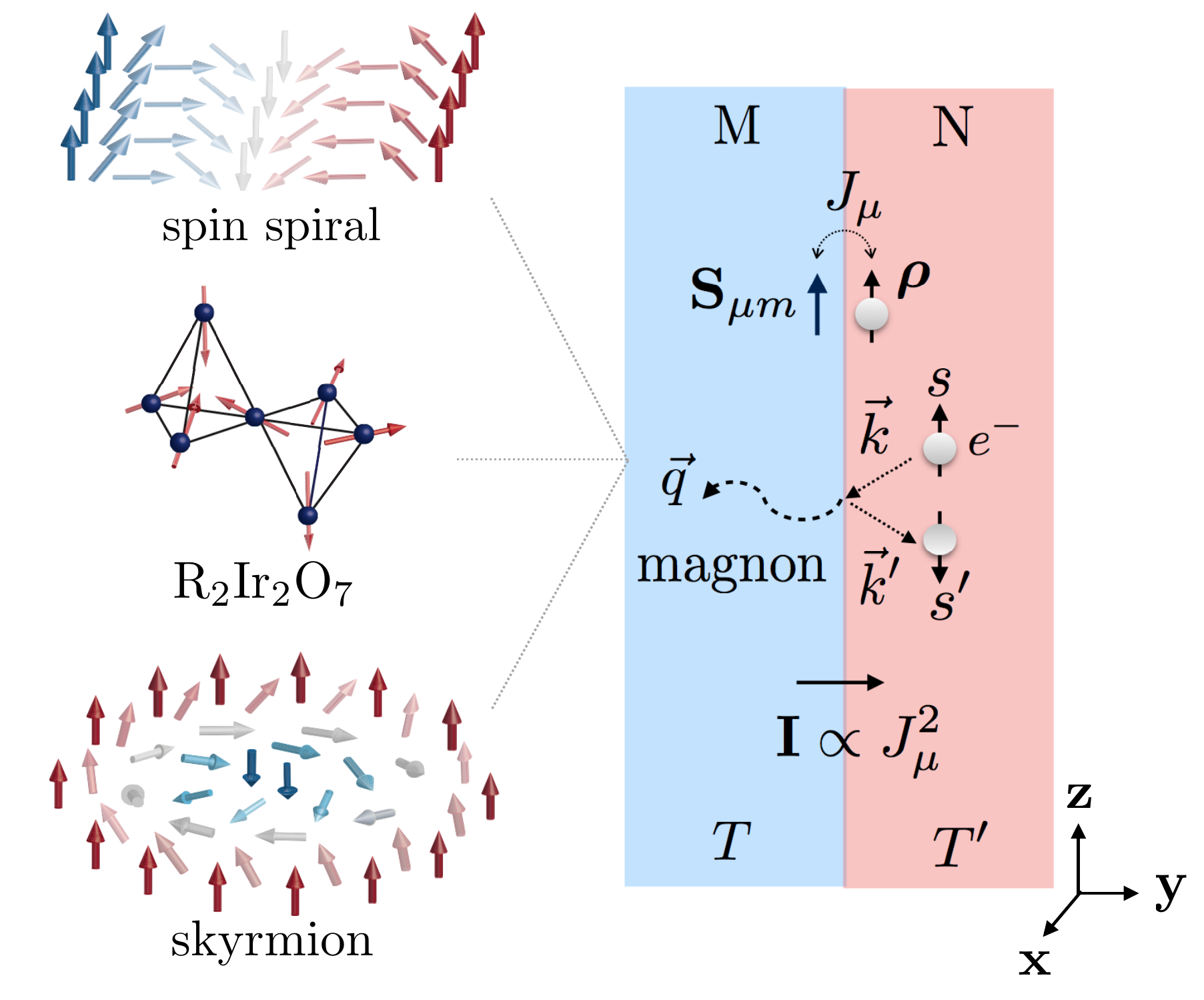}
\caption{Noncollinear magnetic insulator$|$normal metal (M$|$N) heterostructure.  We consider an arbitrary magnetic ordering, e.g., from the top to the bottom left, a spin spiral, the all-in/all-out ordering of R$_2$Ir$_2$O$_7$ pyrochlore iridates or a skyrmion lattice.   The interfacial exchange interaction between the spin $\mathbf{S}_{\mu m}$, located at the site $\mu$ of the $m$th magnetic supercell,  and the electronic spin density $\boldsymbol{\rho}$ is parametrized by the exchange parameter $J_{\mu}$. The coupling engenders inelastic electron-magnon scattering. Namely, a spin-$s$ electron impinging on the interface with wavevector $\vec{k}$  is reflected as a spin-$s'$ electron with wavevector $\vec{k}'$,  generating a magnon with wavevector $\vec{q}$ on the magnetic side.  The difference between the insulator, $T$, and metal, $T'$, temperatures leads to an imbalance between these events and the reciprocal processes, giving rise to a net interfacial spin-current density $\mathbf{I} \propto J^2_{\mu}$. }
\label{Fig1}
\end{figure} 

Heretofore, the SSE has been extensively investigated only for a limited number of insulating systems with collinear magnetic order~[\onlinecite{xiao2010,adachi2010,adachi2011,adachi20112, adachi2013,hoffman2013,Onhuma2013,Rezende2014,Rezende2016,Bender2016,kamra2017}], with a particular focus on the ferrimagnetic yttrium iron garnet (YIG).  
For this ferrimagnet, the SSE has proved to serve as a sensitive probe of magnetization dynamics, allowing one to measure its magnon diffusion~[\onlinecite{cornelissen2015}] and thermalization lengths~[\onlinecite{prakash2018}], as well as the strength of magnetoelastic interactions~[\onlinecite{kikkawa2016}].

Magnetic insulators, however, display a large variety of magnetic orders. Spin-orbit coupling, such as the Dzyaloshinskii-Moriya interaction, favors localized magnetic textures with nontrivial topology such as skyrmions~[\onlinecite{fert2017}]. Strong competing exchange interactions between neighboring magnetic ions represent another source of noncollinearity~[\onlinecite{ramirez1994}]. Moreover, a subtle interplay between these  forces and the electronic interactions is responsible for the metal to noncollinear magnetic-insulator transition  in  pyrochlore iridates~[\onlinecite{gardner2010,krempa,Laurell,Sagayama,Donnerer}].  Contrary to collinear systems, a non-collinear spin texture might engender a field-tunable SSE, as it typically displays a continuous response to a magnetic field through changes in the spin order canting.  Moreover, as  non-collinear orders often result from spin-orbit coupling (SOC), the dependence of the SSE on the orientation of the external field is expected to be more complex than for collinear systems. To properly account for these effects, a microscopic theory  for the heat-to-spin interconversion in a noncollinear magnet|normal metal heterostructure is necessary.

In this work, we investigate the spin current resulting from the interfacial exchange interaction between the normal-metal spin density and the spins of a system with arbitrary magnetic ordering,  as illustrated in Fig.~\ref{Fig1}.  Specifically, we focus on the thermally-activated spin current, i.e., the spin current due to the interfacial SSE, which originates from inelastic magnon-electron scattering at the interface. Starting with equilibrium states of the magnetic and conducting systems in the absence of the interfacial coupling and using the Kubo formula up to the second order in the interfacial exchange parameter, we derive a general expression for the spin current, valid for any magnetic arrangement.  Finally, we check our formula against two well-known collinear magnetic systems, a ferromagnetic and a bipartite antiferromagnetic insulating films and recover established results.

\section{Model}

In this section we consider a noncollinear magnetic  insulator$|$normal metal  heterostructure. We present our model for the magnetic insulator and the normal metal, and we introduce the exchange  Hamiltonian describing the interfacial coupling among them.

\subsection{Magnetic system}
The magnetic system is modeled using a supercell approach, i.e., it consists of a supercell, containing $N$ magnetic sites, that repeats periodically in the $xz$ plane, as depicted in Fig.~\ref{Fig1}.  
%The position of each magnetic site can be decomposed by $\vec{r}_{i}$ =$\vec{r}_{m} + \vec{r}_{\mu}$, where $\vec{r}_{m}$ and $ \vec{r}_{\mu}$ are a primitive and a basis vectors, respectively. 
The initial input of our model  is the classical spin arrangement  of the  magnetic supercell, which can be found analytically or numerically depending on the complexity of the  magnetic Hamiltonian.  
The position of each magnetic site can be written as $\vec{r}_{i}=\vec{r}_{m}+\vec{r}_{\mu}$, where $\vec{r}_m$ is a 2$d$ Bravais lattice vector labelling the $m$th supercell and $\vec{r}_{\mu}$ is the position of the $\mu$th magnetic site   within the supercell, respectively. For each site $\mu$ of the $m$th supercell, we can orient a 
spin-space Cartesian coordinate system such that the $\mathbf{z}$ axis  locally lies along the classical  orientation of the onsite spin operator $\mathbf{S}'_{ \mu m}$. The 
latter can be related to the spin operator $\mathbf{S}_{ \mu m}$ in the global frame of reference via the transformation
\begin{align}
\mathbf{S}_{\mu m}=\mathcal{R}_{z}(\phi_{\mu}) 
\mathcal{R}_{y}(\theta_{\mu})\mathbf{S}'_{\mu m }\,,
\end{align}
 where the matrix $\mathcal{R}_{z (y)}(\zeta)$ describes a right-handed rotation by an angle $\zeta$ about the $\mathbf{z}$ ($\mathbf{y}$) 
axis, and $\theta_{\mu}$ ($\phi_{\mu}$) is the polar (azimuthal) angle of the classical  orientation of the spin $ \mathbf{S}_{ \mu m}$. 

At temperature $T \ll T_{c}$, with $T_{c}$ being the magnetic ordering temperature, we can access the magnon spectrum by linearizing the Holstein-Primakoff transformation in the local frame of reference~[\onlinecite{holstein1940}]: 
\begin{align}
S_{ \mu m}'^{+} =&S_{ \mu m}'^{x} +i S_{ \mu m}'^{y}= \sqrt{2S} b^{ \dagger}_{ \mu m}, \nonumber \\
S_{\mu m }'^{z} =& S -b^{ \dagger}_{ \mu m } b_{ \mu 
m}\,.
\end{align}
Here, $S$ is the classical spin  (in units of $\hbar$) and $b^{ \dagger}_{ \mu m}$ ($b_{ \mu m}$) the magnon creation (annihilation) operator at the site $\vec{r}_{m}$, obeying the boson commutation relation $[ b_{ \mu m}, b^{\dagger}_{ \nu  m' } ]=  \delta_{\mu \nu} \delta_{m m'} $. Next, we truncate the magnetic Hamiltonian beyond the quadratic 
terms in the Holstein-Primakoff bosons and perform a Fourier transform in the $xz$ plane, i.e., 
\begin{align}
b_{ \mu m }=\frac{1}{N_{s}} \sum_{\vec{k}} e^{i \vec{k} \cdot \vec{r}_{m}} 
b_{\vec{k} \mu}\,,
\end{align}
with $\vec{k}$ being a 2$d$ wavevector and $N_{s}$  the number of magnetic supercells. Generally, the resulting Hamiltonian  is not block-diagonal and a Bogoliubov transformation is required to access the spin-wave spectrum $\omega_{\vec{k} \mu}$. The latter can be defined as~[\onlinecite{keffer}],
\begin{align}
b_{\vec{k} \mu}=\sum_{\nu } \left[ M_{\mu \nu}(\vec{k}) c_{\vec{k} \nu} +  N_{\mu \nu}(\vec{k}) c^{ \dagger}_{-\vec{k} \nu } \right]\,,
\end{align}
where the magnetic excitations obey the Bose-Einstein statistics, i.e.,  $\langle  \hat{c}^{\dagger}_{\vec{k} \mu } \hat{c}_{\vec{k}' \nu }  \rangle = n_{B}( \beta \omega_{\vec{k} \mu }) \delta_{\vec{k} \vec{k}'} \delta_{\mu \nu}$, with $n_{B}(x)=(e^{x} -1)^{-1}$, and $M_{\vec{k}}$ and  $N_{\vec{k}}$ are both $N \times N$ matrices. The matrices $M_{\vec{k}}$ and  $N_{\vec{k}}$ are  inputs to our theory and depend on the specific form of the magnetic Hamiltonian.
\subsection{Electronic system}
We treat the conducting side as a spin-degenerate normal metal of thickness $d$. Its spin density (in units of $\hbar$) can be written as
\begin{align}
\boldsymbol{\rho}(\vec{r}_{i})=\frac{1}{2}\sum_{\sigma \sigma' } \psi^{\dagger}_{ \sigma}(\vec{r}_{i}) \boldsymbol{\sigma}_{\sigma \sigma'} \psi_{\sigma'}(\vec{r}_{i})\,,
\label{spindensityelectron}
\end{align}
where $\boldsymbol{\sigma}$ is a vector of Pauli matrices.  For  an electron state with spin $\sigma \hbar/2$ along the $z$-direction (with $\sigma=\pm$), we expand  the corresponding itinerant-electron field operator 
\begin{align}
 \psi_{\sigma }(\vec{r}_{i})=\sum_{\vec{k} \ell} \psi_{ \vec{k} \ell}(\vec{r}_{i}) a_{\vec{k} \ell \sigma}\,,
\end{align}
in terms of electron annihilation operators $a_{\vec{k} \ell \sigma}$ on the basis of Bloch wavefunctions $\psi_{ \vec{k} \ell}(\vec{r}_{i})=e^{i \vec{k} \cdot \vec{r}_{i}} u_{\vec{k} \ell}$. Here,  the quantum number  $\ell$ identifies the electron state in the spin-transport direction (e.g., a quantum-well state). The electronic Hamiltonian is given by
\begin{align}
\mathcal{H}_{e}= \sum_{\vec{k} \ell \sigma} \epsilon_{\vec{k} \ell} a^{\dagger}_{\vec{k} \ell \sigma} a_{\vec{k} \ell \sigma}\,,
\end{align}
with $\langle a^{\dagger}_{\vec{k}' \ell' \sigma'} a_{\vec{k} \ell \sigma} \rangle = n_{F}\left( \beta' \epsilon_{\vec{k} \ell}  \right)  \delta_{\sigma \sigma'} \delta_{\vec{k} \vec{k}'} \delta_{\ell \ell'}$, where $n_{F}(x)=(e^x+1)^{-1}$ is the Fermi distribution function, $\epsilon_{\vec{k} \ell}$ the single-electron energy (measured with respect to the chemical potential) and $T'=\beta'^{-1}$ the common temperature of the electron bath.

\subsection{Interfacial Hamiltonian}

The interfacial exchange interaction between the electronic spin density and the spins of the magnetic system is given by
\begin{align}
\mathcal{H}_{\text{int}}= - \sum_{m \mu} J_{\mu}  \;\boldsymbol{\rho}(\vec{r}_{m}+\vec{r}_{\mu}) \cdot  \mathbf{S}_{\mu m}\,,
\label{tt}
\end{align} 
where $J_{\mu}$ is the exchange-interaction strength at the magnetic site $\mu$, due to the overlap between itinerant and localized orbitals.

Our goal is to determine the spin current driven by a temperature difference between the magnetic and the metallic side. As we will show, the latter arises from thermally-activated electron-magnon scattering at the interface. 
 Thus, while rewriting Eq.~(\ref{tt}) in terms of  second-quantized operators, we  discard terms involving the $z$-component of the spin-density operator (in the local frame), i.e., $ S'^{z}_{ \mu m} = S-b_{ \mu m }^{\dagger} b_{\mu m}$. The first term, i.e., $\propto S a^{\dagger}_{\vec{k} \ell \sigma} a_{\vec{k}' \ell' \sigma}$, describes elastic scattering of electrons off the static spin density of the magnetic system and does not depend on the applied thermal bias~[\onlinecite{Bender2016}]. Within mean field and far from the magnetic ordering temperature, i.e., $N_{\mu}/S \ll 1$, with $N_{\mu}$ being the number of magnons at the site $\mu$, the second term reduces to a correction to the leading elastic-scattering processes, i.e., $\propto N_{\mu} a^{\dagger}_{\vec{k} \ell \sigma} a_{\vec{k}' \ell' \sigma}$. Beyond mean field, the term $\propto  a^{\dagger}_{\vec{k} \ell \sigma} a_{\vec{k}' \ell' \sigma} b_{  \mu m}^{\dagger} b_{  \mu m}$ leads to higher order corrections that we neglect here.
Thus, Eq.~(\ref{tt}) can be rewritten as 
\begin{align}
\mathcal{H}_{\text{int}}=&  \sum_{n=1}^{3} \sum_{\sigma \sigma'} \sum_{\substack{\vec{k} \vec{k}' \vec{q} \\  \vec{G}  \nu \ell\ell'}}  \delta_{\vec{q}+\vec{k}-\vec{k}',\vec{G}} \nonumber \\
\times &\left[ V^{(n)}_{ \nu \ell \ell' }(\vec{k}, \vec{k}', \vec{q}) c^{ \dagger}_{\vec{q} \nu} a^{\dagger}_{\vec{k} \ell \sigma} L^{(n)}_{\sigma \sigma'} a_{\vec{k}' \ell' \sigma'}  +   \text{H.c.} \right]\,,
 \label{109}
\end{align}
where $\vec{G}$ is a reciprocal lattice vector~[\onlinecite{footnoteBravais}], $L=\{\sigma^{+}, \sigma^{-}, \sigma_{z}\}/2$, and 
\begin{align}
V^{ (n)}_{\ \nu \ell \ell' }(\vec{k}, \vec{k}', \vec{q})&= -\sqrt{2S} \sum_{\mu} J_{\mu} u^{*}_{\vec{k} \ell}(\vec{r}_{\mu}) u_{\vec{k}' \ell'}(\vec{r}_{\mu}) e^{i (\vec{k}'-\vec{k}) \cdot \vec{r}_{\mu}} \nonumber \\
& \times \left[ f_{n \mu} M^{\dagger}_{\mu \nu}(\vec{q})  + g_{n \mu} N_{\mu \nu}(-\vec{q}) \right]\,, 
\end{align}
with  $f_{1 \mu}=g^{*}_{2\mu}=\sin^2 (\theta_{\mu}/2) e^{-i \phi_{\mu}} $, $g_{1 \mu}=f^{*}_{2 \mu}=\cos^2 (\theta_{\mu}/2) e^{-i \phi_{\mu}}$ and $f_{3\mu}=g_{3\mu}= - \sin\theta_{\mu}/2 $. \\

\section{Results.}

In this section we discuss our analytical result for the interfacial spin current, and we check it against two well-known collinear magnetic systems.

\subsection{Interfacial spin current}

The conservation of spin density on the conducting side allows us to define the interfacial spin-current density operator (per unit of area) into the normal metal as
\begin{align}
\mathbf{i}=d \hbar \dot{\boldsymbol{\rho}}= - i  d  \left[ \boldsymbol{\rho}, \mathcal{H}_{\text{int}} \right]\,.
\end{align}
 Using the Kubo formula to second order in the exchange parameter $J_{\mu}$~[\onlinecite{bruus}], we find
the $\alpha$-polarized spin-current density as
\begin{align}
&I_{\alpha}= \frac{1}{(2\pi)^2 A}  \sum_{\substack{\vec{k} \vec{k'} \vec{q}  \\ \vec{G} \nu \ell \ell'}}  
\text{Re}\bigg[ \eta V^{ (\beta)}_{ \nu \ell \ell'}(\vec{k}, \vec{k}',\vec{q}) V^{* (\gamma)}_{ \nu \ell \ell'}(\vec{k}, \vec{k}',\vec{q}) \nonumber \\
&- \epsilon V^{ (\delta)}_{ \nu \ell \ell'}(\vec{k}, \vec{k}',\vec{q}) V_{  \nu \ell \ell'}^{ * (\zeta)}(\vec{k}, \vec{k}',\vec{q})  \bigg]  B_{\nu \ell \ell'}(\vec{k}, \vec{k}' , \vec{q})  \delta_{\vec{q}+\vec{k}-\vec{k}',\vec{G}}\,,
\label{current}
\end{align}
with $A$ being the interfacial area and
\begin{align}
&B_{\nu \ell \ell'}(\vec{k}, \vec{k}' , \vec{q})=    \int d\omega_{1} \;
d\epsilon_{1} \; d\epsilon_{2} \; \delta (\epsilon_{1} - \epsilon_{2} +\omega_{1}) \mathcal{A}_{\nu}(\vec{q},\omega_{1})  \nonumber \\& \mathcal{A}_{\ell}(\vec{k}, \epsilon_{1}) \mathcal{A}_{\ell'}(\vec{k}', \epsilon_{2}) \bigg\{ n_{B}(\beta \omega_{1}) n_{F}(\beta' \epsilon_{1}) \left[1-n_{F}(\beta' \epsilon_{2}) \right] \nonumber \\&   -\left[ 1 +n_{B}(\beta \omega_{1}) \right]  \left[1-n_{F}(\beta' \epsilon_{1}) \right] n_{F}(\beta' \epsilon_{2})    \bigg\}\,.
\label{spectral}
\end{align} 
Here, $\mathcal{A}_{\nu}(\vec{q},\omega)$ and $\mathcal{A}_{\ell}(\vec{k},\epsilon)$ are, respectively, the magnon and electron spectral functions.
For $\alpha=z$, we have $\beta,\gamma=1$, $\delta,\zeta=2$ and $\eta,\epsilon=1$. For $\alpha=x \; (y)$, we have $\gamma=1$, $\beta,\delta=3$,  $\zeta=2$ and $\eta,\epsilon=4 \; (\eta, -\epsilon=4i)$.

The first term on the right-hand side of Eq.~(\ref{spectral}) describes an electron scattering inelastically off the interface  and creating a magnon in the insulator, while the second term accounts for the reciprocal process.   The difference between the insulator, $T$, and metal, $T'$, temperatures breaks the balance between these two processes, driving a net spin-current density across the interface. 
Equation~(\ref{current})  shows that, as a consequence of the translational invariance in the $xz$ plane, the inelastic  electron-magnon scattering at the interface conserves the linear 2$d$ momentum, modulo the reciprocal lattice basis (i.e., umklapp processes). On the other side,  as the noncollinearity of the magnetic system breaks spin-rotational invariance, no spin component is conserved at the interface. Thus, inelastic scattering events without electronic spin-flip contribute as well to the total spin current.

Given any spin arrangement, Eq.~(\ref{current}) allows one to calculate the interfacial spin current, driven by an arbitrary temperature difference, as function of material-specific microscopic parameters.
By neglecting vertex corrections, disorder in the normal metal can be accounted for via the electronic spectral function. For the magnetic side, instead, the effects of disorder can be included by  increasing the size of the magnetic supercell such that it greatly exceeds the  primitive one.

\subsection{Collinear magnetic systems}In order to test our results, we consider  two collinear magnetic systems, i.e., a ferromagnetic and a bipartite antiferromagnetic thin films.  For simplicity, we treat both the electronic and magnetic systems as clean and non-interacting, i.e., $\mathcal{A}(\vec{q},\omega_{1})=2\pi \delta (\omega_{1}-\omega_{ \vec{q}})$ and $\mathcal{A}(\vec{k},\epsilon)=2\pi \delta (\epsilon-\epsilon_{\vec{k}})$, and we neglect umklapp scattering. Moreover, we assume that 
the common electronic temperature $T'$ and the single electron energy $\epsilon_{\vec{k}}$ are both much smaller than the Fermi energy $\epsilon_{F}$. Thus, we can treat the electron density  of states as a constant, i.e., $D$. \\
First, we consider a ferromagnetic system with collinear ground state defined by $\mathbf{S} \parallel - \mathbf{z}$~[\onlinecite{footnoteisotropicity}], i.e., its magnon modes carry spin $\hbar$ along the $z$ direction. We assume that the corresponding Hamiltonian can be diagonalized in terms of a standard Bogoliubov transformation~[\onlinecite{White}], i.e., 
\begin{align}
\begin{bmatrix}
 b_{\vec{k}} \\ b^{\dagger}_{-\vec{k}}\end{bmatrix}= \begin{bmatrix}  u_{\vec{k}} & v^{*}_{\vec{k}} \\ v_{\vec{k}} & u_{\vec{k}} \end{bmatrix} \begin{bmatrix} c_{\vec{k}} \\ c^{\dagger}_{-\vec{k}}
\end{bmatrix}\,,
\label{easyBG}
\end{align}
with $|u_{\vec{k}}|^2-|v_{\vec{k}}|^2=1$. Generally, magnons flow in the direction opposite to a thermal bias, i.e.,  from the hot to the cold region~[\onlinecite{Bender2012}]. Thus, if the magnetic-insulator  temperature  is higher than the normal-metal one, i.e.,  $T>T'$, we expect the $z$-polarized spin-current density to flow into the conducting side, i.e., $I_{z} >0$. This is correctly predicted by Eq.~(\ref{current}), which reduces to
\begin{align}
I_{z} = \frac{4\pi S(D J)^2}{A} \sum_{\vec{q}} V_{\vec{q}} \; \omega_{\vec{q}} \left[n_{B}(\beta \omega_{ \vec{q}})- n_{B}(\beta'  \omega_{ \vec{q}})  \right]\,,
\label{526}
\end{align}
with 
\begin{align}
V_{\vec{q}}=(1/AD)^2\sum_{\vec{k} \vec{k}'} \delta (\epsilon_{F} -\epsilon_{\vec{k}}) \delta(\epsilon_{F}+\omega_{\vec{q}}-\epsilon_{\vec{k'}}) \delta_{\vec{k} + \vec{q},\vec{k}' }\,.
\end{align}

Secondly, we consider a bipartite antiferromagnet with collinear ground state given by $\mathbf{S}_{1(2)} \parallel \pm \mathbf{z}$, with $1 (2)$ labelling the sublattice site. The corresponding Hamiltonian can be recasted in diagonal form by defining~[\onlinecite{Rezende2016}]
\begin{align}
\begin{bmatrix}
 b_{1 \vec{k}} \\ b^{\dagger}_{2 -\vec{k}}\end{bmatrix}= \begin{bmatrix}  u_{\vec{k}} & -v_{\vec{k}} \\ -v_{\vec{k}} & u_{\vec{k}} \end{bmatrix} \begin{bmatrix} c_{1 \vec{k}} \\ c^{\dagger}_{2 -\vec{k}}
\end{bmatrix}\,,
\label{easyBG}
\end{align}
where $c^{\dagger}_{1 (2)}$ is the creation operator of a magnon mode carrying spin angular momentum $\mp \hbar$ and energy $\omega_{1 (2)}$. 
From Eq.~(\ref{current}), we find the  $z$-polarized spin-current density  as~[\onlinecite{footnoteisotropicity}]
\begin{align}
I_{z}&=\frac{S}{2 \pi^2 A^3}\sum_{\vec{k} \vec{k}' \vec{q} }  \bigg[  B_{2}(\vec{k}, \vec{k}' , \vec{q})
 |J_{1} v_{\vec{q}} \; e^{i \vec{q} \cdot \vec{r}_{1}} +J_{2} u_{\vec{q}} \; e^{i  \vec{q} \cdot \vec{r}_{2}} |^2  \nonumber \\
&- B_{1}(\vec{k}, \vec{k}' , \vec{q}) |J_{1} u_{\vec{q}} \; e^{i \vec{q} \cdot \vec{r}_{1}}+J_{2} v_{\vec{q}} \; e^{i  \vec{q} \cdot \vec{r}_{2}} |^2 \bigg]  \delta_{\vec{k} + \vec{q},\vec{k}' }\,,
\label{166}
 \end{align}
 where $B_{1(2)}$ is given by Eq.~(\ref{spectral}).
Equation~\eqref{166} shows that cross-sublattice terms contribute to the spin-current density. However, electron interference effects  vanish when the two antiferromagnetic modes are equally populated. For degenerate dispersions, Eq.~(\ref{166}) indeed becomes
\begin{align}
I_{z} = \frac{4\pi SD^2 (J^2_{2}-J^2_{1})}{A} \sum_{\vec{q}} V_{\vec{q}} \; \omega_{\vec{q}} \left[n_{B}(\beta \omega_{ \vec{q}})- n_{B}(\beta'  \omega_{ \vec{q}})  \right].
\label{540}
\end{align}
\begin{figure}[t!]
\includegraphics[width=0.75\linewidth]{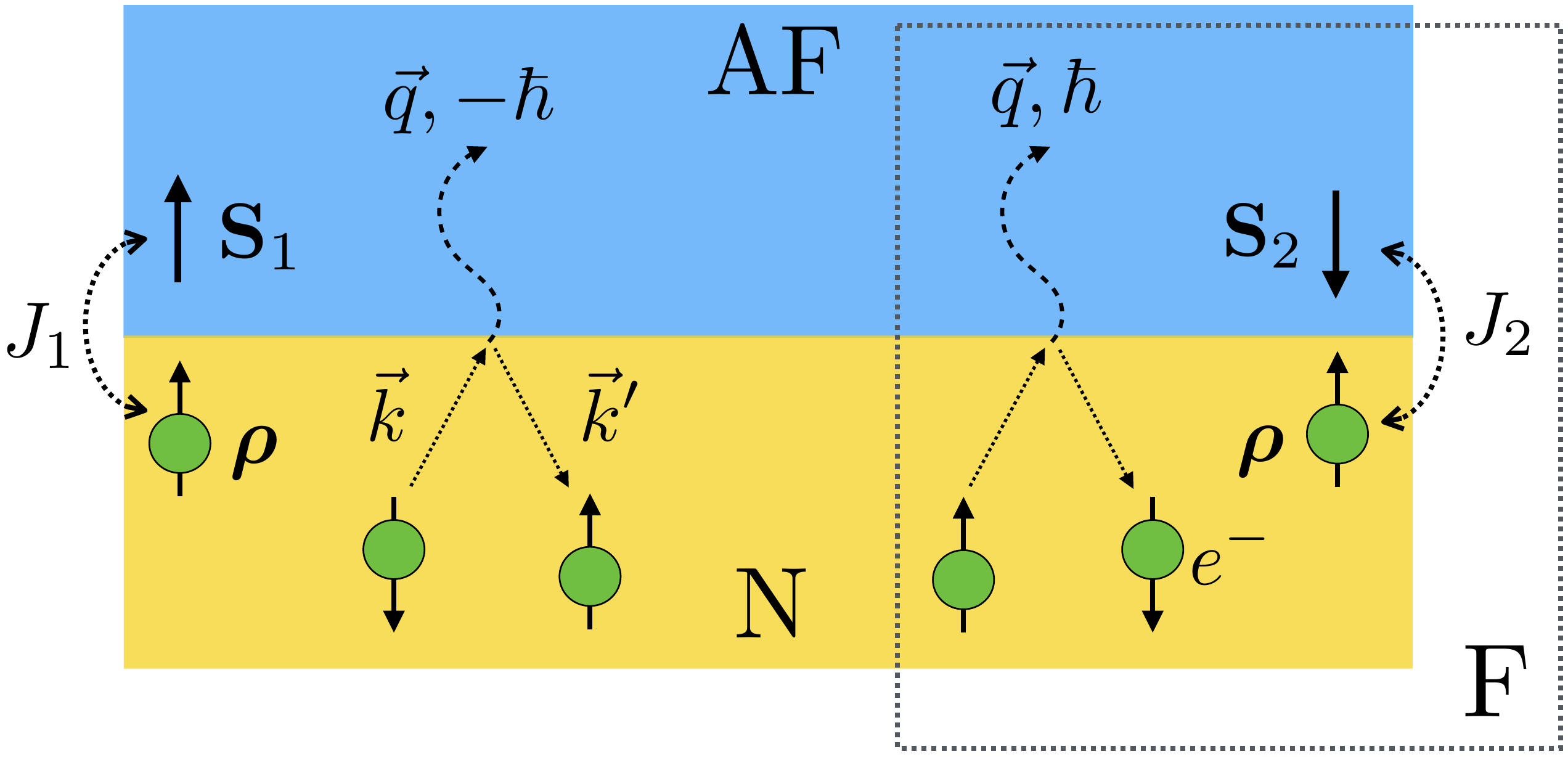}
\caption{Antiferromagnetic insulator (AF) $|$ normal metal (N) heterostructure. The conductor spin density $\boldsymbol{\rho}$ couples to the spin $\mathbf{S}_{1 (2)} $ of the antiferromagnetic sublattice $1$($2$) with exchange strength $J_{1(2)}$. Due to the collinearity of the magnetic system, the spin component parallel to the magnetic symmetry axis must be conserved. Thus, the only processes that are allowed include a spin-down (up) electron can scatter inelastically off the interface, flipping its spin and creating a magnon carrying angular momentum $\mp \hbar$, and their reverse. The dashed line encompasses the allowed scattering processes in the ferromagnetic (F) system we consider. } 
\label{figure2}
\end{figure} 
Equation~(\ref{540}), which agrees with the results of Ref.~[\onlinecite{kamra2017}], offers a simple physical interpretation: as magnons associated with the sublattice 2 (1) carry spin angular momentum $\pm \hbar$,  when the metallic side is hotter,  they  lead to a positive (negative) contribution to the $z$-polarized spin current.  In a collinear magnetic system, the inelastic magnon-electron processes driving the spin current must conserve the component of the spin parallel to the magnetic symmetry axis, i.e., the $\mathbf{z}$ axis. Thus, an electron with spin-up (down) impinging on the interface is reflected with flipped spin, creating a magnon with angular momentum  $\pm \hbar$, as illustrated in Fig.~\ref{figure2}.
Equation~(\ref{540}) shows that, even when the antiferromagnetic modes are degenerate, the interfacial spin-current density (\ref{540}) is finite if the symmetry between antiferromagnetic sublattices at the interface is broken, i.e., $J_{1} \neq J_{2}$, which confirms the predictions of Ref.~[\onlinecite{bender2017}]. 
\\
\section{Discussion and conclusions}
In this work, we address  thermally-driven spin transport at a noncollinear magnetic insulator $|$ normal metal interface, deriving an expression for the spin current applicable to any magnetic ordering. Our theory reproduces correctly the interfacial spin transport in collinear magnets $|$ normal metal heterostructures.

For thin films, along with bulk systems with strong spin-orbit coupling, the interface can represent the bottleneck for spin transport. 
In this scenario, our result captures the main contribution to the thermally-driven spin current. Otherwise, our expression for the spin current serves as a boundary condition, which should be complemented by an appropriate theory for bulk magnon transport~[\onlinecite{Bender2016}].
The interfacial spin density induced by the spin current can be measured on the conducting side of the interface via the magneto-optical Kerr effect~[\onlinecite{Choi2014}].  
Instead, to relate our results for the spin current  to a measurable ISHE voltage, the conductor has to be modelled as a bulk system and our theory  complemented  with transport equations describing spin diffusion on the metallic side~[\onlinecite{Cornelissen2016}].

While here we focus on the spin current due to the SSE, our framework could be easily extended to include the spin pumping and spin-transfer torque contributions to the overall interfacial spin current, as well as additional interfacial interactions~[\onlinecite{Amin2016}]. 
Future work should investigate the  corrections arising from a non-perturbative treatment of the interfacial exchange interaction.

\section{ACKNOWLEDGEMENTS}
The authors thank S.A. Bender and  B. Ma  for insightful discussions. B.F. was supported by the Dutch Science Foundation (NWO) through a Rubicon grant and Y.T. by NSF under Grant No. DMR-1742928. G. A. Fiete has been supported by the NSF under grants NSF DMR-1729588,  NSF Materials Research Science
and Engineering Center Grant No. DMR-1720595, and a Simons Fellowship.


\begin{thebibliography}{99}

\bibitem{wakeham} N. Wakeham, A. F. Bangura, X. Xu, J.-F. Mercure, M. Greenblatt and N. E. Hussey,  Nat. Commun. \textbf{2}, 396 (2011).

\bibitem{mitali} M. Banerjee, M. Heiblum, V. Umansky, D. E. Feldman, Y. Oreg, and  A. Stern, Nature \textbf{559}, 205 (2018). 

\bibitem{spinliquid} M. Ye, G. B. Hal\'asz, L. Savary, and L. Balents,
Phys. Rev. Lett. \textbf{121}, 147201  (2018).

\bibitem{Bauer2010} G. E. W. Bauer, A. H.  MacDonald, and S.  Maekawa,  Solid State Commun. \textbf{150}, 459–460 (2010); G. E. W. Bauer, E. Saitoh, and B. J. van Wees, Nat. Mat. \textbf{11}, 391 (2012).

\bibitem{Zutic2004} I. Zutic, J. Fabian, and S. Das Sarma,  Rev. Mod. Phys.
\textbf{76}, 323 (2004).

\bibitem{uchida2008}  K. Uchida, J. Xiao, H. Adachi, J. Ohe, S. Takahashi, J. Ieda, T. Ota, Y. Kajiwara, H. Umezawa, H. Kawai, G. E. W. Bauer, S. Maekawa, and E. Saitoh, Nat. Mat. \textbf{9}, 894–897 (2010).

\bibitem{Uchida2014} K. Uchida, M. Ishida, T. Kikkawa, A. Kirihara, T. Murakami, and E. Saitoh, J. Phys.: Condens. Matter \textbf{26}, 343202 (2014).


\bibitem{Saitoh2006} E. Saitoh, M. Ueda, H. Miyajima, and G. Tatara, Appl. Phys. Lett. \textbf{88}, 182509 (2006).

\bibitem{Kimura2007} T. Kimura, Y. Otani, T. Sato, S. Takahashi, and S. Maekawa, Phys. Rev. 
Lett. \textbf{98}, 156601 (2007). 

\bibitem{Miao} B. F.  Miao, S. Y. Huang, D. Qu, and C. L. Chien, 
Phys. Rev. Lett. \textbf{111}(6), 066602 (2013).

\bibitem{Sinova2015} J. Sinova, S. O. Valenzuela, J. Wunderlich, C. H. Back, and T. Jungwirth, Rev. Mod. Phys. \textbf{87}, 1213  (2015).



\bibitem{xiao2010} J. Xiao, G. E. W. Bauer, K. Uchida, E. Saitoh, and S. Maekawa, Phys. Rev. B \textbf{81}, 214418 (2010).


\bibitem{adachi2010} H. Adachi, K. Uchida, E. Saitoh, J. Ohe, S. Takahashi, and S. Maekawa, Appl. Phys. Lett. \textbf{97}, 252506 (2010). 

\bibitem{adachi2011} H. Adachi, J. Ohe, S. Takahashi, and S. Maekawa,
Phys. Rev. B \textbf{83}, 094410 (2011).

\bibitem{adachi20112} J. Ohe, H. Adachi, S. Takahashi, and S. Maekawa, Phys. Rev. B \textbf{83}, 115118 (2011). 

\bibitem{adachi2013} H. Adachi, K. Uchida, E. Saitoh, and S. Maekawa, Rep. Prog. Phys. \textbf{76}, 036501 (2013).

\bibitem{hoffman2013} S. Hoffman, K. Sato, and Y. Tserkovnyak,  Phys. Rev. B \textbf{88}, 064408 (2013).


\bibitem{Onhuma2013} Y. Ohnuma, H. Adachi, E. Saitoh, and S. Maekawa
Phys. Rev. B \textbf{87}, 014423 (2013).

\bibitem{Rezende2014}S. M. Rezende, R. L. Rodr\'iguez-Suárez, R. O. Cunha, A. R. Rodrigues, F. L. A. Machado, G. A. Fonseca Guerra, J. C. Lopez Ortiz, and A. Azevedo
Phys. Rev. B \textbf{89}, 014416 (2014).

\bibitem{Rezende2016} S. M. Rezende, R. L. Rodriguez-Suarez, and A. Azevedo, Phys. Rev. B \textbf{93}, 014425 (2016).

\bibitem{Bender2016} S. A. Bender and Y. Tserkovnyak, Phys. Rev. B \textbf{91}, 140402(R) (2015).

\bibitem{kamra2017} A. Kamra and W. Belzig, Phys. Rev. Lett. \textbf{119}, 197201 (2017).



\bibitem{cornelissen2015} L. J. Cornelissen, J. Liu, R. A. Duine, J. Ben Youssef, and B. J. van Wees, Nat. Phys. \textbf{11}, 1022 (2015).

\bibitem{prakash2018} A. Prakash, B. Flebus, J. Brangham, F. Yang, Y. Tserkovnyak, and J. P. Heremans, Phys. Rev. B \textbf{97}, 020408(R) (2018).



\bibitem{kikkawa2016} T. Kikkawa, K. Shen, B. Flebus, R. A. Duine, K. Uchida, Z. Qiu, G.
E. W. Bauer and E. Saitoh, Phys. Rev. Lett. \textbf{117}, 207203 (2016); B. Flebus, K. Shen, T. Kikkawa, K. Uchida, Z. Qiu, E. Saitoh, R. A. Duine, and G. E. W. Bauer,  Phys. Rev. B \textbf{95}, 144420 (2017); L. J. Cornelissen, K. Oyanagi, T. Kikkawa, Z. Qiu, T. Kuschel, G. E. W. Bauer, B. J. van Wees, and E. Saitoh,
Phys. Rev. B \textbf{96}, 104441.

\bibitem{fert2017} A. Fert, N. Reyren, and V. Cros, Nat. Rev. Mat. \textbf{2}, 17031 (2017).

\bibitem{ramirez1994} A. P. Ramirez,  Ann. Rev. Mater. Sci. \textbf{24}, 453 (1994).

\bibitem{gardner2010} J. S. Gardner, M. J. P. Gingras, and J. E. Greedan, Rev. Mod. Phys. \textbf{82}, 53 (2010).


\bibitem{krempa} W. Witczak-Krempa, G. Chen, Y. B. Kim, and L. Balents, Annu. Rev. Condens. Matter Phys. \textbf{5}, 1 (2014).
57.


\bibitem{Laurell} P. Laurell and G. A. Fiete,
Phys. Rev. Lett. \textbf{118}, 177201 (2017).

\bibitem{Sagayama} H. Sagayama, D. Uematsu, T. Arima, K. Sugimoto, J. J. Ishikawa, E. O'Farrell, and S. Nakatsuji,
Phys. Rev. B \textbf{87}, 100403(R) (2013).

\bibitem{Donnerer} C. Donnerer, 
M. C. Rahn, 
M. Moretti Sala, 
J. G. Vale,
 D. Pincini, J. Strempfer, M. Krisch, 
D. Prabhakaran, 
A. T.  Boothroyd, and 
D. F. McMorrow,
Phys. Rev. Lett. \textbf{117}, 037201 (2016).

\bibitem{holstein1940} T. Holstein and H. Primakoff, Phys. Rev. \textbf{58}, 1098 (1940).

\bibitem{keffer} F. Keffer, \textit{Spin Waves} (Springer, Verlag, 1966), Vol. XVI- IIP, Handbuch der Physik.



\bibitem{footnoteBravais} Supposing a lattice-matched heterostructure, we assume that the magnetic and conducting sides share the same 2$d$ (i.e., defined within the $xz$ plane) Bravais lattice.


\bibitem{bruus} H. Bruus, and K. Flensberg, \textit{Many-Body Quantum Theory in Condensed Matter Physics: An Introduction} (Oxford University Press, Oxford, 2004).
\bibitem{footnoteisotropicity}{Since the interfacial exchange interaction~\eqref{tt} is isotropic, we  find  analogous results for the $x$- and $y$-polarized spin current when we assume the order parameter to lie along, respectively, the -$x$ and -$y$ direction. }

\bibitem{White} R. M. White, \textit{Quantum Theory of Magnetism} (McGraw-Hill, New York, 1970).


\bibitem{Bender2012} S. A. Bender, R. A. Duine, and Y. Tserkovnyak,
Phys. Rev. Lett. \textbf{108}, 246601 (2012).



\bibitem{bender2017} S. Bender, H. Skarsv$\dot{\text{a}}$g, A. Brataas, and R. A. Duine, Phys. Rev. Lett. \textbf{119}, 056804 (2017).

\bibitem{Choi2014} G.-M. Choi, B.-C. Min, K.-J. Lee, and D. G. Cahill,  Nat. Commun. \textbf{5}, 4334 (2014); G.-M. Choi,  C.-H. Moon, B.-C. Min, K.-J. Lee, and D.G. Cahill, Nat. Phys. \textbf{11}, 576 (2015).

\bibitem{Cornelissen2016} Y. Tserkovnyak and S. A. Bender, Phys. Rev. B \textbf{90}, 014428 (2014);  L. J. Cornelissen, K. J. H. Peters, G. E. W. Bauer, R. A. Duine, and B. J. van Wees, Phys. Rev. B \textbf{94}, 014412 (2016).

\bibitem{Amin2016} V. P. Amin, and M. D. Stiles, Phys. Rev. B \textbf{94}, 104419 (2016).



\end{thebibliography}
\end{document}